# Microelectronic interconnects based on carbon nanotubes

Franz Kreupl[*], Andrew P. Graham, Maik Liebau, Georg S. Duesberg,
Robert Seidel and Eugen Unger
Infineon Technologies, Corporate Research, Otto-Hahn-Ring 6, D-81739 Munich, Germany
[*] Phone: +49 89 234 44618, Fax: +49 89 234 9552068, e-mail: franz.kreupl@infineon.com

**Abstract**

Carbon nanotubes have emerged as a possible new material for electronic applications. They show promising characteristics for transistors as well as for interconnects. Here we review their basic properties and focus on the status of nanotubes with respect to their application as interconnects and discuss the challenges facing their integration.

**Introduction**

In Figure 1(a) the hexagonal structure of a single atomic layer of graphite (graphene) is shown, which can be wrapped up to form a single-walled carbon nanotube (SWCNT), as shown in Fig. 1(b) [1]. When several nanotubes are nested concentrically inside one another a multi-walled carbon nanotube (MWCNT) is obtained (Fig.1(c)). The diameter d of SWCNTs is in the range of 0.4 nm to 5 nm and for MWCNTs typically between 4 nm and 100 nm. A tight-binding calculation of the energy diagram of a graphene layer is shown in Fig. 1(f) and demonstrates that graphite is a semi-metal, as the valence and conduction band touch each other exactly at the Fermi point. A closer look in Fig. 1(g) reveals that the energy diagram in the vicinity of the

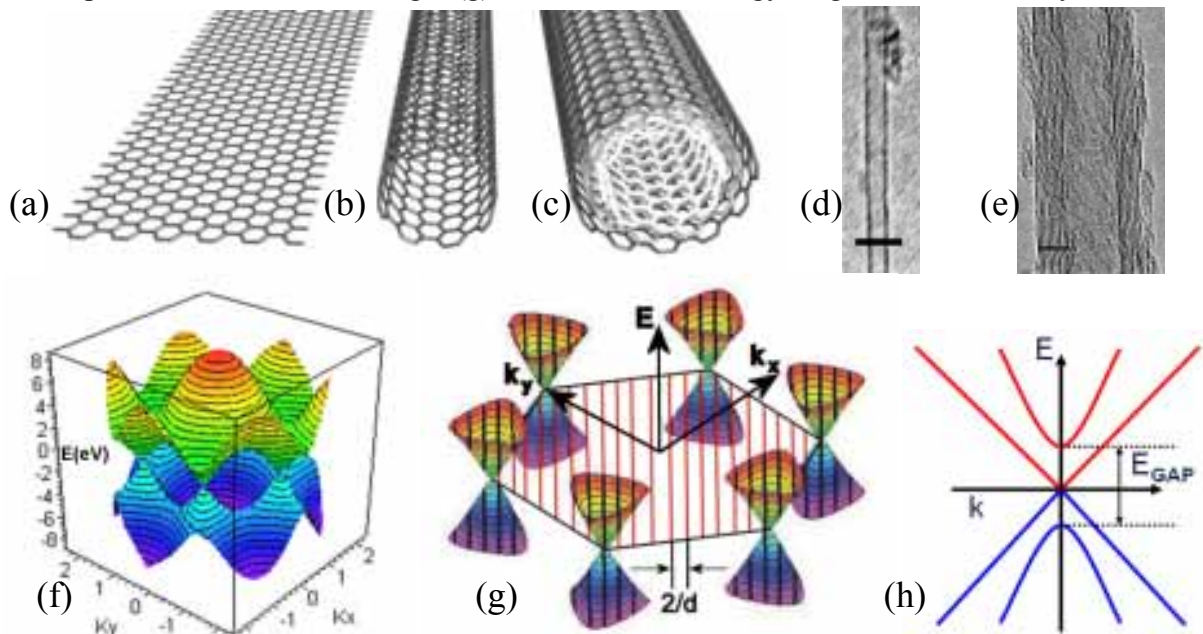

**Figure 1**. (a) A monolayer of graphite. (b) A single-walled nanotube. (c) A multi-walled nanotube. (d) TEM image of a SWCNT (scale bar 2 nm). (e) TEM image of a MWCNT (scale bar 2 nm). (f) Band diagram of a graphene layer showing the hexagonal symmetry. (g) Band diagram in the vicinity of the Fermi energy. The energy landscape is cut into slices with distance 2/d as indicated by the thick lines, if the quantization around the circumference of the nanotube is imposed. (h) Depending on whether the cut hits the centre of the cone or not, the SWCNT is metallic or semiconducting with energy gap $E_{gap}$.

Fermi energy resembles hexagonally arranged cones. By imposing the quantization condition, $\mathbf{C_h \cdot k} = 2\pi q$, of the wave vector k around the circumference $C_h$, the energy landscape is cut into slices of allowed states as indicated in Fig. 1(g) by the thick solid lines. The slices are separated by 2/d, with d being the diameter of the tube. Depending on how the graphene sheet is wrapped up to make the nanotube, the cut goes through the center of the cone and the resulting nanotube is metallic, or if not, the nanotube will be semiconducting (See Figure 1(h)).

Over the past several years detailed information on the electrical transport behaviour of SWCNTs has been obtained [2]. The deduced mean free path ($l_{mfp}$) of 1- 10 µm for the charge carriers can be attributed to the reduced phase space accessible for scattered charge carriers in 1-dimensional systems. In ordinary metals the mean free path is of the order of 10 nm due to phonon scattering. Despite the low density of states in the nanotube (as compared to metals) the specific resistivity of nanotubes is of the order of µΩcm, i.e. comparable to or even better than the bulk value of copper. The band structure of a metallic (7,7) single nanotube with a diameter around 1 nm is shown in Figure 2(a). The electrons have almost a linear massless dispersion in the vicinity of the Fermi energy and a density of states of $2 \cdot 10^7/(eV \cdot cm)$.

Under ambient conditions CNTs are usually p-doped, which shifts the Fermi level to negative values. This has no significant effect for SWCNTs with small diameters, because the additional states are far away, but can contribute to the conductivity of MWCNTs with larger diameters, as additional states become accessible [3]. This is indicated in Figure 2(c), where the band structure of a 20 nm diameter tube has been calculated. Due to the light p-doping of the tube the Fermi energy is located in the valence band giving rise to more conducting channels (See Figure 2(d)). The doping is mainly caused by charge transfer from the surrounding material and can also be due to adsorbed water [3]. The resistance of a single-walled nanotube can be approximated by R= $R_C$ +h/(4$e^2$)+ h/(4$e^2$)(L/$l_{mfp}$), where $R_C$ is the contact resistance at both ends of the tube with length L. For ideal contacts $R_C \approx 0$ and the tubes behave ballistically for lengths below 1-10 µm. Therefore, the preferred application of nanotubes is in the range of their ballistic mean free path. The resistance of MWCNTs is less well understood, because on one hand more states are available in one single shell due to the wider diameter, whereas on the other hand the contributions to the conductance due to the other shells needs to be considered. Reported values for MWCNT resistances range between 0.35 kΩ and 20 kΩ. Measured current densities in nanotubes exceed values of $10^9$ A/$cm^2$, which is almost a factor of 1000 higher than copper can withstand. The high current carrying capabilities can be understood from the strong covalent bonds in this material as compared to the metallic bonds in ordinary metals. Looking at the future requirements for interconnects in the ITRS roadmap nanotubes may seem to be an attractive candidate for future interconnect material because current densities exceeding the possible values for copper are required [4,6].

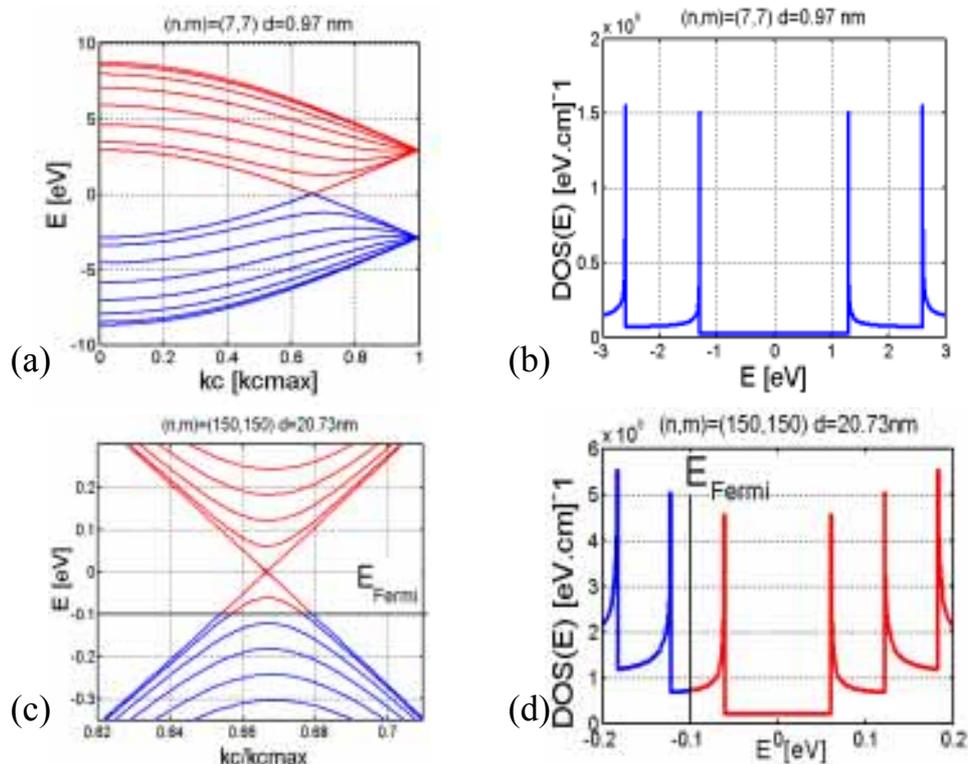

**Figure 2.** (a) Band structure of a (7,7) single-walled nanotube with approximately 1 nm diameter and the resulting density of states (b) in the vicinity of the Fermi level (E= 0 eV). (c) Band structure of the outer shell of a (150,150) MWCNT with approximately 20 nm diameter. The p-doped shell reveals more density of states (d) for conduction and has therefore a lower resistance than a SWCNT.

**Carbon Nanotube Transistors**

The semiconducting nanotubes can be used to build very efficient field-effect transistors. CNT devices are therefore under consideration for future nano-devices, which will outperform silicon devices once their manufacturability is solved [2,4]. Meanwhile our group has discovered that CNTs can also be used to build very efficient power switches and *the first nanotube power transistor* delivering milliampere currents has been presented [5]. The prototype, shown in Figure 3, is already able to operate LEDs and electro-motors and can be easily scaled to higher loads. Since one nanotube can only carry about 24 µA, about 300 CNTs have been connected in parallel to switch 2.4 mA at $V_{ds}$= 1 V, as shown in Fig. 3(b)-3(d).

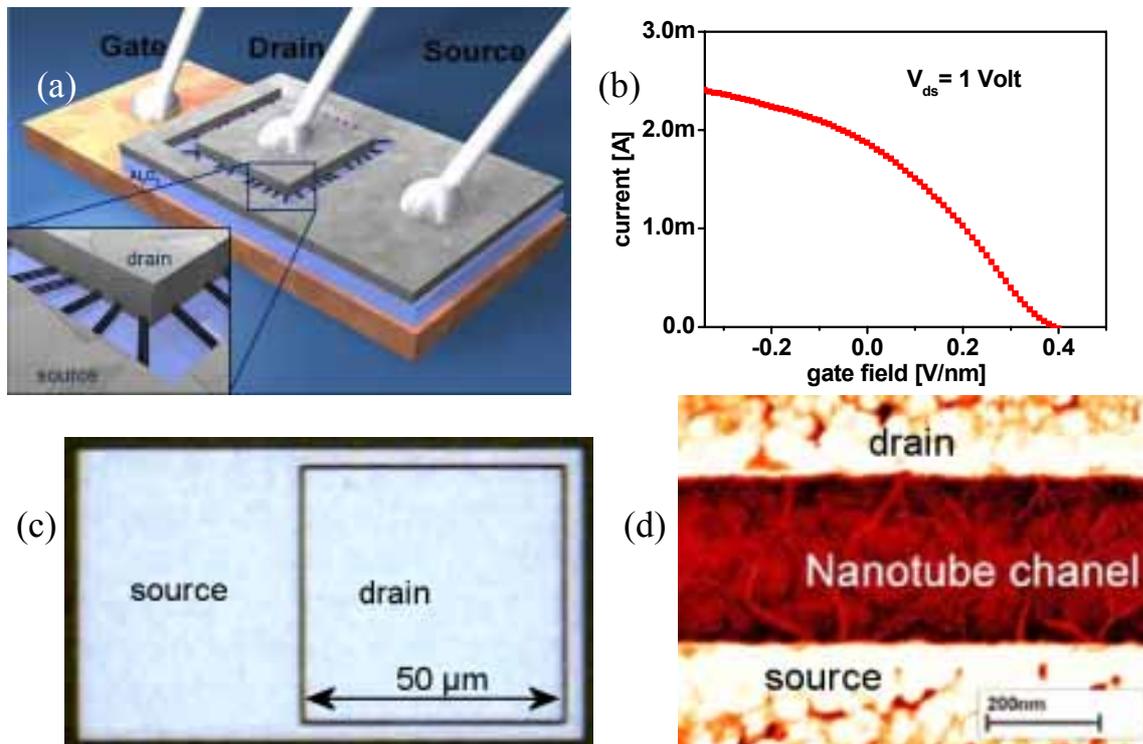

**Figure 3.** (a) Schematics of the nanotube power transistor prototype with about 300 nanotubes in parallel [5]. (b) Output characteristic demonstrating current switching of 2.4 mA at 1 Volt. (c) Optical image of the device with 200 µm nanotube channel width. (d) SEM picture of the nanotube channel.

**Carbon nanotubes for future interconnects**

The quasi ballistic transport properties and high current carrying capacity makes CNTs ideal candidates for interconnects. As a substitute for copper or tungsten, CNTs have to compete with both the resistance *and* current carrying capacity of the metals. Whereas the measured current densities outperform any known material, the length dependence of the resistance needs to be investigated further. Recent work has focused on the length dependence of SWCNTs and an average resistance of 4-6 kΩ/µm length has been evaluated for length scales above 1 µm [2,7,8]. Figure 4(a) summarizes the measured trends for lengths up to about 10 mm [7]. The deviation from pure ballistic behaviour starts at 500 - 1000 nm. The quantum resistance of ~ 6.4 kΩ also sets a lower limit to the lowest resistance, even if the conductor length is reduced by half. For ordinary metals like copper or tungsten this is considered to happen only below the free mean path of ~ 40 nm. The consequences for the specific resistivity are detailed in Figure 4(b). For length scales exceeding 10µm, SWCNTs are almost one order of magnitude better than any known metal or conductors. For shorter lengths below ~ 500 nm the specific resistivity deteriorates drastically, which is due to the huge free mean path of ~1 µm. Therefore, CNTs are especially suited for short ( ~1µm) high aspect structures, where the size-effect dominates the resistivity of metals [9], or for very long interconnects, made of densely packed CNTs. Due to the big challenges of manipulating and handling CNTs efficiently, the most viable way to approach the use of CNTs is the implementation as vias and contact holes. First integration schemes have been developed by three groups to address this specific problem [6,11,12].

**Vias and contact holes for future interconnects**

The ITRS roadmap details the requirements for specific resistivity and current density of the individual technology nodes. One of the major concerns is the electro-migration limit in copper, which leads to voids and subsequent failure. However, the requested current density in the ITRS is only given for the intermediate wiring levels. The constraints for local vias or contacts are even more stringent as the same current has to pass through an even narrower via which has the minimum pitch. The situation is depicted in Fig. 5(a) where the required current densities are recalculated for local vias based on the information detailed in the ITRS year 2003 roadmap. It is clear that vias below the 45 nm node are even more prone to failure by electro-migration than the 90 nm node vias [10]. The required current densities are at the limit that ordinary copper can withstand. Therefore, CNTs have been proposed to replace copper in future vertical interconnects [5]. Two different schemes for via filling with CNTs are shown in Fig. 5 together with an example of a 400 nm diameter via filled with an array of MWCNTs.

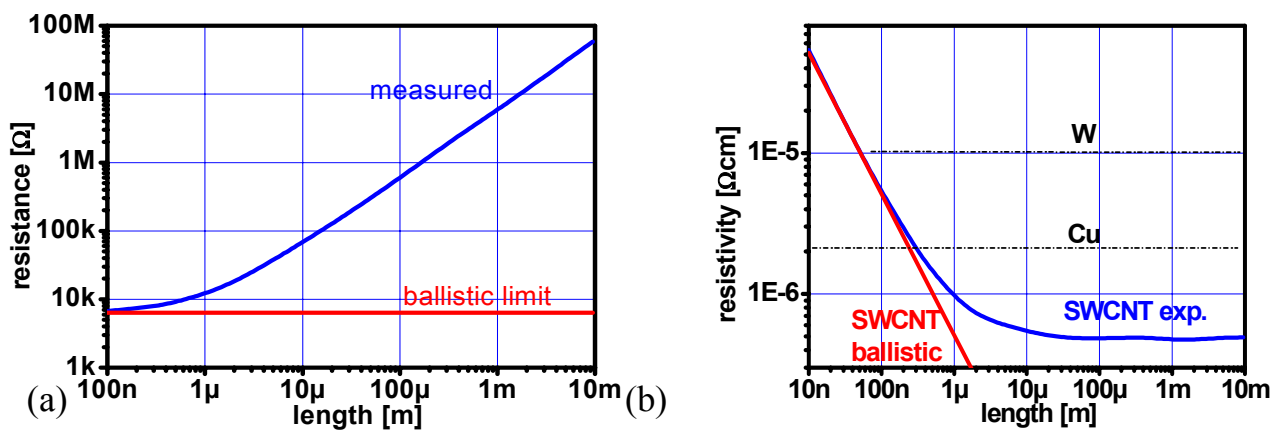

**Figure 4.** (a) Length dependence of the resistance of SWCNTs deduced from measurements and compared to the ballistic limit (adapted from [7,8]). (b) The specific resistivity of SWCNTs compared to bulk values of Cu and W (neglecting size-effects [9]).

Small vias (~ 20 nm ⌀) can be filled with one single MWCNT consisting of several individual shells that all contribute to the conductance. An alternative approach would be to implement a densely packed array of SWCNTs or even small diameter MWCNTs. The minimal achievable resistance is strongly correlated with the ability to make very densely packed arrays of the order of $1/nm^2$. This can be seen in Fig. 5 (b) where a calculation for varying densities has been performed. A comparison with copper is not possible as no data for copper wires with 20 nm diameter are available in the moment and the size effect in metals will severely contribute in this regime [9].

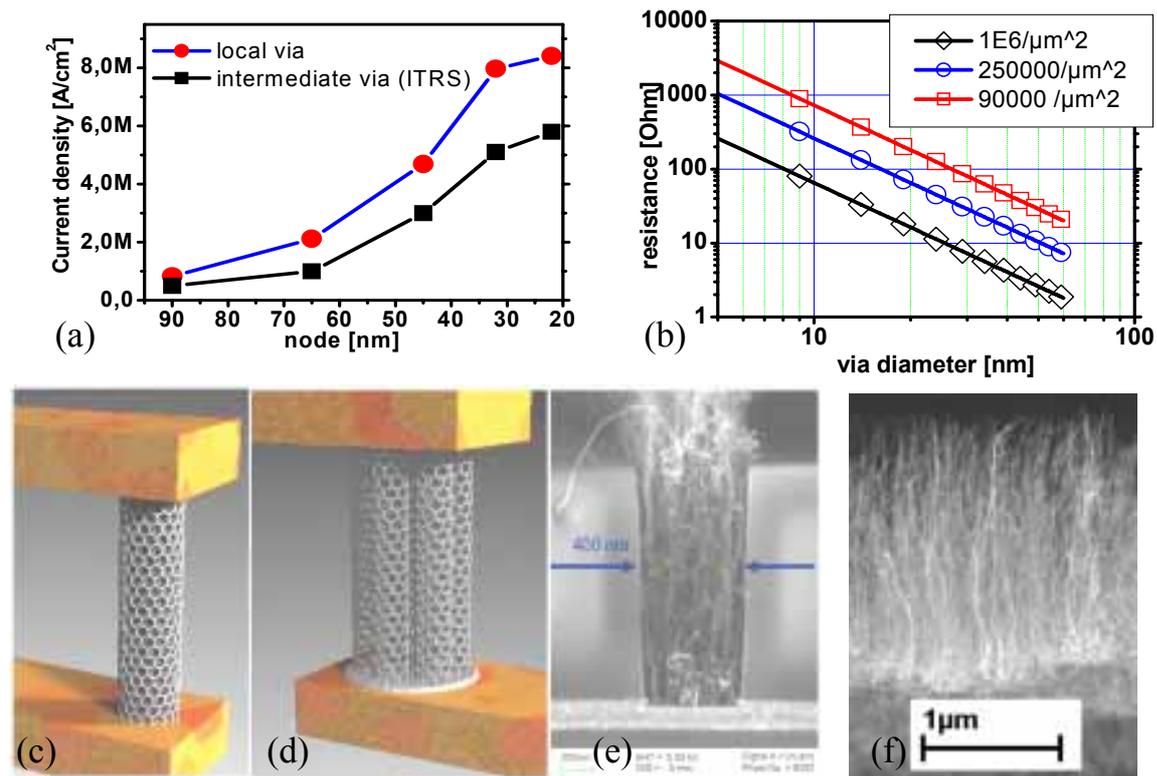

**Figure 5.** (a) Maximum current density as given by ITRS roadmap and for local vias designed at minimum pitch. (b) Calculated via resistance for SWCNT arrays of varying densities. Square vias and a resistance of 6.5 kΩ per tube are assumed. (c) Via or contact hole made from one SWCNT or one MWCNT. (d) Via or contact hole made from densely packed array of SWCNTs or MWCNTs. (e) Example of a 400 nm wide via filled with an array of MWCNTs. (f) SEM image of SWCNT growth with a density of ~ $100000/\mu m^2$.

**Multi-walled carbon nanotube via chain for the 22 nm node**

For the 22 nm mode we succeeded in making vias with an aspect ratio of about 10 by e-beam lithography and subsequent dry etching [13,14]. A buried catalyst approach was used, where the catalyst resides on top of the lower metal layer and the etch process stops on this catalyst layer [15]. MWCNTs have been grown in the vias by chemical vapour deposition [6,16]. The individual MWCNTs fill the whole 20 nm wide via, as shown in Fig. 7. After deposition of top contacts a via chain consisting of two MWCNT vias in series have been measured with a current density exceeding $10^7$ A/cm$^2$. The average resistance for one via is, in this specific case, between 20 k$\Omega$ and 30 k$\Omega$, which is still too high for low ohmic contacts. The high resistance may be caused by the bottom and top contacts and it has been shown that it could be lowered considerably by appropriate annealing steps. Furthermore, it is imperative to make sure that all shells of the MWCNT are properly contacted and can contribute to the conductance [15]. Whereas the achieved current densities exceed already the values for metals, the integration challenges to achieve an almost two orders of magnitude lower resistance are considerable.

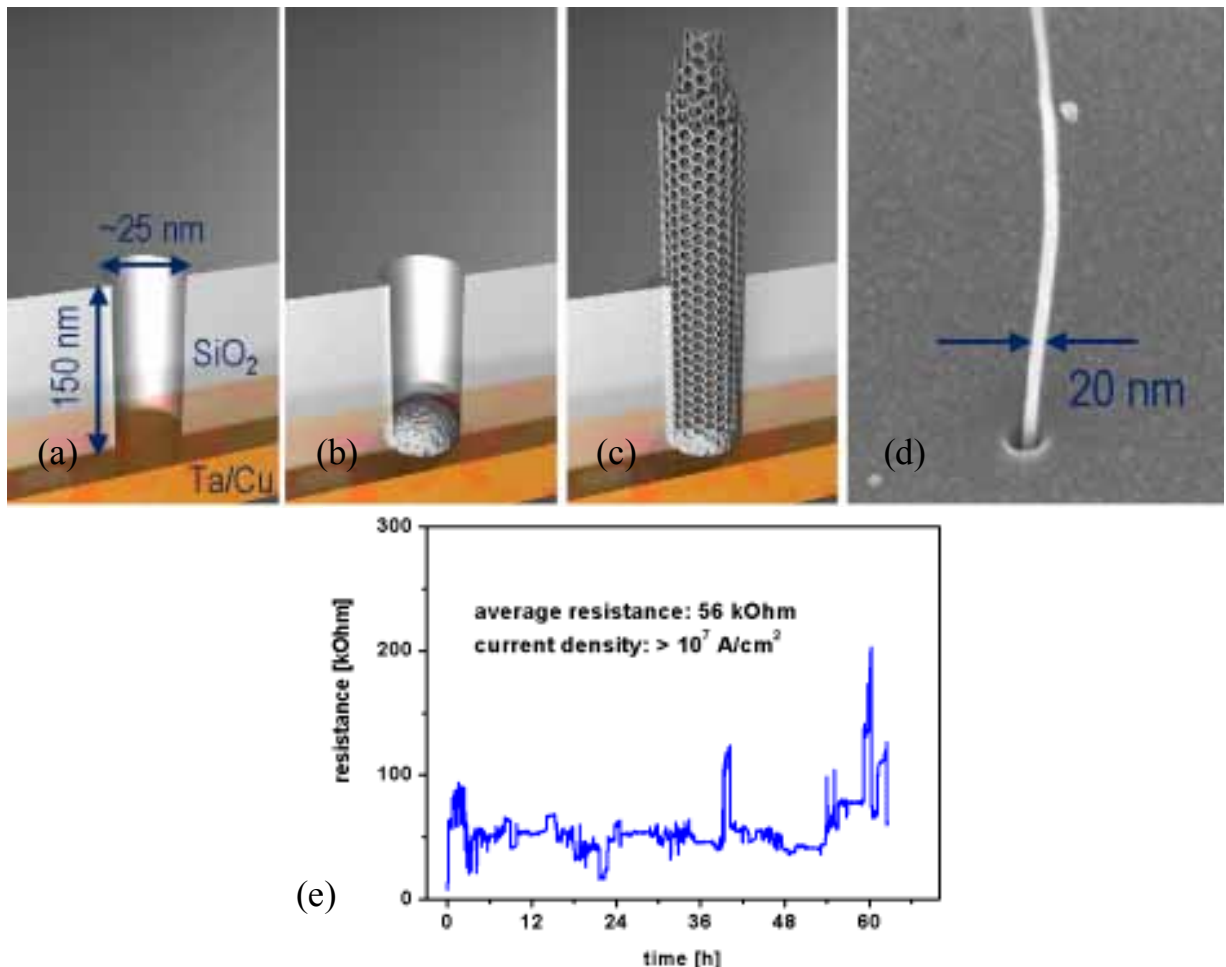

**Figure 7.** Fabrication process for a 22 nm node via. (a) Etching of the via structure with stop on the catalyst layer. (b) Catalyst formation during heating. (c) Growth of a MWCNT in the via. (d) A MWCNT protrudes from the via hole, which is contacted again with a top contact. (e) Reliability measurement in a via chain consisting of two MWCNT vias in series.

**Acknowledgement**

We would like to thank W. Hoenlein for continuous support, W. Pamler for artwork and Z. Gabric for expert technical assistance. The work is supported by BMBF under contract # 13N8402.